# ASYMPTOTES OF THE NONLINEAR TRANSFER AND THE SWELL SPECTRUM IN THE FRAME OF THE KINETIC EQUATION

**Vladislav G. Polnikov**[1†1], **Fangli Qiao**[2,3] **and Yong Teng**[2]

[1]A.M. Obukhov Institute of Atmospheric Physics of RAS, Moscow, Russia
[2]First Institute of Oceanography of SOA, Qingdao, China
[3]Laboratory for Regional Oceanography and Numerical Modeling, Qingdao National Laboratory for Marine Science and Technology, Qingdao, China

**Abstract**

The kinetic equation for a gravity wave spectrum is solved numerically to study the high frequencies asymptotes for the one-dimensional nonlinear energy transfer and the variability of spectrum parameters that accompany the long-term evolution of nonlinear swell. The cases of initial two-dimensional spectra $S(\omega,\theta)$ with the frequency decay-law $S(\omega) \sim \omega^{-n}$ (for $n = 6, 5, 4$ and 3.5) and various initial functions of the angular distribution are considered. It is shown that at the first step of the kinetic equation solution, the nonlinear energy transfer asymptote has the power-like decay-law, $Nl(\omega) \sim \omega^{-p}$, with values $p \leq n-1$, valid for cases in which $n \geq 5$, and the difference, $n-p$, changes significantly when $n$ approaches 4. On time scales of evolution greater than several hundred initial wave-periods, in every case, a self-similar spectra $S_{sf}(\omega,\theta)$ is established with the frequency decay-law of form $S(\omega) \sim \omega^{-4}$. Herein, the asymptote of nonlinear energy transfer becomes negative in value and decreases according to the same law (i.e., $Nl(\omega) \sim -\omega^{-4}$). The peak frequency of the spectrum, $\omega_p$, migrates to the low-frequency region such that the angular and frequency characteristics of the two-dimensional spectrum $S_{sf}(\omega,\theta)$ remain constant. However, these characteristics depend on the degree of angular anisotropy of the initial spectrum. The solutions obtained are interpreted, and their connection with the analytical solutions of the kinetic equation by Zakharov and co-authors for gravity waves in water is discussed.

## 1. Introduction

In the study of properties for the random field of nonlinear gravity waves in water, an important role is played by the four-wave kinetic equation (KE) of the form

$$\frac{\partial N(\mathbf{k}_0)}{\partial t} = 4\pi \int M^2_{0,1,2,3} N3_{0+1-2-3} \delta_{\mathbf{q},0+1-2-3} d\mathbf{k}_1 d\mathbf{k}_2 d\mathbf{k}_3 \ , \tag{1a}$$

---

[1] † Correspondent author is Vladislav Polnikov: polnikov@mail.ru

where $N(\mathbf{k}_i)$ is the wave-action spectrum in the wave-vector **k**-space ($i$=0,1,2,3), $M^2_{0,1,2,3}$ is the second power of the matrix elements corresponding to the four-wave nonlinear interactions, $N3_{0+1-2-3}$ is the cubic-power functional of $N(\mathbf{k}_i)$ with the form

$$N3_{0+1-2-3} \equiv N(\mathbf{k}_2)N(\mathbf{k}_3)[N(\mathbf{k}_0)+N(\mathbf{k}_1)] - N(\mathbf{k}_0)N(\mathbf{k}_1)[N(\mathbf{k}_2)+N(\mathbf{k}_3)] \quad (1b)$$

and

$$\delta_{\mathbf{q},0+1-2-3} \equiv \delta(\omega_0+\omega_1-\omega_2-\omega_3)\delta(\mathbf{k}_0+\mathbf{k}_1-\mathbf{k}_2-\mathbf{k}_3) \quad (1c)$$

is the Dirac's delta-function responsible for a resonant feature of the four-wave interactions. The relationship between wave vector $\mathbf{k} = (k_x, k_y) = (k, \theta)$ and wave frequency $\omega$ is given by the dispersion relation. For gravity waves, it is as $\omega(\mathbf{k}) \equiv \omega(k) = (gk)^{1/2}$, where $g$ is the gravity acceleration. The wave energy spectrum, $S(\omega,\theta)$, given in the frequency-angular space, $(\omega,\theta)$, and related to $N(\mathbf{k})$ by the ratio

$$S(\omega,\theta) \propto \omega N(\omega,\theta) \propto (\omega^4/g^2)N(\mathbf{k}) \quad , \quad (2)$$

is dealt with below.

The expressions for matrix elements $M_{0,1,2,3}$ are well known (Hasselmann 1962; Zakharov 1968; Badulin et al. 2005) but extremely cumbersome, so they are not given here. It is important to note only that the kinetic integral (KI), standing in the r.h.s of (1a), formally preserves the total wave energy, $E = \iint S(\omega,\theta)d\omega d\theta$, the total wave action, $N = \iint (S(\omega,\theta)/\omega)d\omega d\theta$, and the total wave momentum, $\mathbf{M} = \iint (\mathbf{k}S(\omega,\theta)/\omega)d\omega d\theta$, under the condition of uniform convergence of the KI (see Hasselmann 1962; Badulin et al. 2005; Geojaev & Zakharov 2017). However, the calculations of KI show that simultaneous conservation of these integrals is not always fulfilled numerically. Therefore, during the solving of KE (1), the conserved quantity may be chosen as an alternative (see below).

Equation (1) was first derived by Hasselmann (1962) in the potential approximation, starting from the Euler's equations for waves in water. The derivation of KE (1) was then carried out by Zakharov (1968) using the technique of the Hamiltonian formalism. The KE has since become an independent subject of study (e.g., Webb 1978; Masuda 1980; Hasselmann & Hasselmann 1981; Polnikov 1989, 1990, 2007; Resio & Perrie 1991; Pushkarev et al. 2003; Badulin et al. 2005; Young & Van Vledder 2003; and references of this paper).

For the case of an angular isotropic spectrum spread throughout the infinite frequency band ($0 < \omega < \infty$), Zakharov and Filonenko (1966) found the analytic solution of KE (1), $S_Z(\omega)$, which puts the KI to zero (i.e., $I_{NL}[S_Z(\omega)] = 0$). The solution of $S_Z(\omega)$ takes the form

$$S_Z(\omega) \propto \omega^{-4}. \tag{3}$$

Zakharov and Zaslavskii (1982) later established that KE (1) has the second analogous solution of the form

$$S_{ZZ}(\omega) \propto \omega^{-11/3}. \tag{4}$$

They have proposed to interpret solutions (3) and (4) as the Kolmogorov-type spectra of the constant energy-flux, $P_E$, directed upward in frequencies, and constant wave-action flux $P_N$, downward in frequencies, respectively. In this interpretation, spectra (3) and (4) acquire the proper Kolmogorov-type representation

$$S_Z(\omega) = c_1 P_E^{1/3} g^{4/3} \omega^{-4}, \quad \text{and} \quad S_{ZZ}(\omega) = c_2 P_N^{1/3} g \omega^{-11/3}, \tag{5}$$

where $c_1$ and $c_2$ are the dimensionless Kolmogorov's constants. It was assumed that the sources and sinks of these fluxes are located at the zero and infinity frequency-points, depending on the case (Zakharov & Zaslavskii 1982). It should be noted, however, that this interpretation is a hypothesis, because it is based only on dimensional considerations and does not follow from the Euler's equations. Later, in a large series of papers by Zakharov and co-authors, the mentioned analytical results were sophisticated in different directions: the generalisation for a case of anisotropic spectra (Zakharov & Pushkarev 1999); the description of asymptote for the spatial and temporal evolution of the self-similar spectra of forms (5) (Pushkarev et al. 2003; Badulin et al. 2005); and the searching convergence conditions for KI (Geojaev & Zakharov 2017; Zakharov 2017). These analytical results are well represented in the cited papers and in numerous references therein.

In parallel, the study of KE was carried out numerically, beginning with the works by Webb (1978), Masuda (1980), and Hasselmann and Hasselmann (1981), devoted to the development of methods for calculation of KI. These methods were further developed by Polnikov (1989), Resio and Perrie (1991), Lavrenov (2004) and van Vledder (2006). Here it should be noted that the most important part of the KI-calculation algorithm is the method of estimating the contribution of integrable singularities in the integrand (the contribution of 'singular points'), resulting from integration of the frequency delta-function (Masuda 1980; Polnikov 1989). The different approaches represent the main differences in the algorithms mentioned above. Polnikov (1989) showed numerically that the contribution of singular points determines the zero-balance for the quantities $E$, $N$ and **M** and the accuracy of the KI-estimate as a whole.

For the first time, a detailed numerical study of the properties for KI of form (1) was carried out by Masuda (1980). It was later expanded by Polnikov (1989) via the introduction of a proper classification of two-dimensional spectra forms $S(\omega,\theta)$. In the later study by Polnikov (1990), the numerical solution of KE (1) was obtained, and it was shown for the first time that the self-similar shape of wave spectrum, $S_{sf}(\omega,\theta)$, is established during the long-term evolution as a result of nonlinear interactions among waves. This result was later confirmed in a series of papers by Zakharov's group (see references in Pushkarev et al. 2003; Badulin et al. 2005) and others (i.e., Lavrenov 2004; Garganier-Renou & Benoit 2007). In addition to Polnikov (1990), other authors (Pushkarev et al. 2003; Badulin et al. 2005) have established that, under the condition of the conservation of wave action $N$, the tail of the self-similar spectrum $S_{sf}(\omega)$ falls according to the law $S_{sf}(\omega,\theta) \sim \omega^{-4}$, and that the shape of the spectral peak for $S_{sf}(\omega,\theta)$ is close in the form of the JONSWAP spectrum, with $\gamma = 3.3$ (Badulin et al. 2005). At the same time, other essential features of the spectrum-peak shape and the process of forming the two-dimensional self-similar spectrum, $S_{sf}(\omega,\theta)$, are still not described, because the main attention was paid to establishment of the Kolmogorov's spectra of forms (3) and (4).

Polnikov (1994) first showed numerically that the spectra of forms (3) and (4) are actually formed as the result of the numerical solution for the generalised KE having the form

$$\partial S(\omega,\theta)/\partial t = I_{NL}[S(\omega,\theta)] + In(\omega,\theta) - Dis(\omega,\theta). \tag{6}$$

In (6), $In(\omega,\theta)$ and $Dis(\omega,\theta)$ are the source and sink functions, respectively, which are separated in frequency space, to ensure the presence of the inertial interval, which is the main condition for applicability of the Kolmogorov's theory (Monin & Yaglom 1971). Despite the technical limitations in performing the calculations, the results of Polnikov (1994) gave grounds for the validity of the hypotheses by Zakharov and Zaslavskii (1982). Similar results were later obtained in numerous studies (e.g., Pushkarev et al. 2003; Lavrenov 2004; Badulin et al. 2005) in which other algorithms were used to compute KI and thicker numerical grids. They confirmed the fact of establishing spectra of forms (3) and (4), depending on the configuration of the source and sink location on the calculating frequency band. In addition, Pushkarev et al. (2003) and Badulin et al. (2005) analysed the long-term asymptote of the peak-frequency downshift, $\omega_p(t)$, estimated the Kolmogorov's constant values, and determined the rate of wave-energy leakage $E(t)$, assuming that the value of total wave action $N$ remains constant.

Nevertheless, studying the property of the KE of form (1) requires its continuation. Namely, the asymptote of nonlinear energy transfer (NLT) at high frequencies ($\omega \gg \omega_p$), both at the initial state and on a long-term scale of the spectrum evolution, is yet to be described, and the quantitative characteristics of the two-dimensional self-similar spectrum shape, $S_{sf}(\omega,\theta)$, as well as their dependence on the initial conditions, are not given. There is also no analysis of the effects of conservation conditions for total wave energy $E$ or wave action $N$ on the shape of the self-similar spectrum, which is realised in the course of numerical solution of KE without sources and sinks (i.e., KE of form [1]).

This paper is devoted to studying these details of numerical solution for KE (1).

## 2. Method of research

In the numerical study of KI and features for the KE-solution, the choice of a certain frequency-angular numerical grid $(\omega,\theta)$ provides the accuracy of calculations (Polnikov 1999; van Vledder 2006). In this study, to examine the asymptote of NLT, an extended frequency grid with an enhanced angular resolution is chosen. It is given by the ratios:

$$0.64 \leq \omega \leq 80 \text{ rad/s} \quad \text{and} \quad -180^\circ \leq \theta \leq 180^\circ; \quad (7a)$$

with

$$\omega_i = \omega_1 q^{i-1} \quad \text{and } \omega_1 = 0.64 \text{ rad/s}, q = 1.05, 1 \leq i \leq I = 100, \text{ and } \Delta\theta = 5^\circ. \quad (7b)$$

The initial high-frequency asymptote of NLT is determined from the results of calculating the KI at the first time-step of the KE solution.

To study the asymptote of NLT and spectral shape resulting from numerical solutions of KE on a long-term scale for spectrum evolution, we used the rarer numerical grid:

$$0.64 \leq \omega \leq 7 \text{ rad/s with } I = 50, \Delta\theta = 10^0, \quad (8)$$

and the previous values for $\omega_1$ and $q$. This was done for reasons of technical difficulty (high time costs and increasing numerical instability) during numerous and long-term calculations on grid (7).

Comparison of the KI estimates on grids (7) and (8) showed an estimate of KI-calculation errors of within 3% to 5%, which is quite acceptable for our purposes, because the errors in the KI-calculations (at each time-step) are accumulated insignificantly during the KE solution up to approximately 1000 steps (corresponding to time scales of the order of $10^5$ units of the initial period, $2\pi/\omega_p(0)$). This previously established feature of the KE solution is due to the property of the KI to smooth spectral irregularities and support a certain shape of the spectrum in the process of solving the KE (Hasselmann & Hasselmann 1981; Polnikov 1990, 1994; van Vledder 2006).

In this paper, all calculations were performed in the dimensional units. Analysis of the forms of numerical solutions of the KE was carried out for evolution times exceeding $10^4/\omega_p(0)$.

The initial wave spectrum was given in the well-known JONSWAP form (J-spectrum)

$$S_J(\omega,\theta,n,\gamma) = S_{PM}(\omega,n)\gamma^{\{(\omega/\omega_p - 1)^2/2\sigma^2\}} \Psi(\omega,\theta) \quad (9)$$

where $\sigma = 0.07$ to $0.09$ is the peak-width parameter of the J-spectrum, $\gamma = 1$ to $7$ is the 'peakness' parameter, $\Psi(\omega,\theta)$ is the frequency-angular form, and

$$S_{PM}(\omega,n) = g^2(\omega^{-n}/\omega_p^{5-n}) exp\left[-(n/4)(\omega_p/\omega)^4\right] \quad (10)$$

is the form of the Pearson-Moskowitz spectrum, generalised to the case of an arbitrary degree $n$ of the spectrum-tail decay (Polnikov 1989). The initial peak frequency, $\omega_p(0)$, was assumed to be 2 rad/s, whilst the initial form of function $\Psi(\omega,\theta)$ was assumed to be independent of the frequency and was given in the form

$$\Psi(\omega,\theta) \equiv \Psi(\theta) = cos^a(\theta/m) \ . \tag{11}$$

Variation of the parameters $a$ and $m$ ensured the assignment of any degree of angular anisotropy of the spectrum used for calculation of the KI at the initial stage of the spectrum evolution.

The algorithm for calculating the KI was used according to work by Polnikov (1989); its high level of quality was shown in the smoothness and numerical stability of the NLT-estimates. Herewith, for the reason of the approximate fulfilment of the conservation of total energy $E$ or wave action $N$ in the course of the numerical solution of the KE (Zakharov 2017), we used two versions of the numerical algorithm, corresponding to the exact conservation of either wave-energy $E$ or wave-action $N$. These versions were realised by applying an exact balance between the positive and negative values for nonlinear transfer for either wave energy or wave action. This approach allows studying the influence of these two evolution regimes (algorithm versions) on the long-term features of KE-solution (see below).

The behaviour of the high-frequency asymptote for the one-dimensional NLT, $NL(\omega)$, was determined by the least-squares estimation of the decay parameter $p$ in the power-law dependence $NL(\omega)$ of the form

$$\int I_{NL}\{S(\omega,\theta,t)\}d\theta \equiv Nl(\omega,t) = const(t) \cdot \omega^{-p} \ . \tag{12}$$

The estimation of $p$ was executed in frequency band $5\omega_p < \omega < (15\text{-}20)\omega_p$ for numerical grid (7).

The numerical solution of KE (1) was performed according to the algorithm by Polnikov (1990), which includes an explicit numerical scheme of the first order of accuracy, linear interpolation of the spectrum from the nodes of the computational grid to the resonance points of a four-wave quadruplet, and a choice of the time step, $\Delta t$, according to the condition

$$\Delta t = 0.03 / min(abs[I_{NL}\{S(\omega,\theta,t)\}/S(\omega,\theta,t)]) \ . \tag{13}$$

Ratio (13) indicates restriction of the spectrum-intensity increase to 3% or less of the current value $S(\omega,\theta,t)$ at each time step. Although this choice of $\Delta t$ led to a significant increase in computing time, it was adopted because it generated good results and ensured the smoothness and numerical stability of the solution.

Following to Polnikov (1989, 1990), the quantitative characteristics of the two-dimensional spectrum-shape were determined in the integral representation by estimating the spectral frequency-width parameter, $B$, and the angular-narrowness function of spectrum, $A(\omega)$, given by the formulas

$$B = E / \omega_p S_p \quad \text{and} \quad A(\omega) = S(\omega, \theta_p)/S(\omega). \tag{14}$$

Here, $S_p = S(\omega_p)$ is the value of the one-dimensional spectrum at the peak frequency, and $\theta_p$ is the direction of propagation of the peak-components for a two-dimensional wave spectrum. Hereafter, $\theta_p = 0$.

## 3. Calculation results and analysis

*3a. Asymptotes of the NLT.*

Calculations of KI on grid (7) were performed for several spectral forms, the representative set of which is presented in Table 1, where the estimates of parameter *p* for the power asymptote of the NLT of form (12) are also given. For purposes of explanation, in figures 1(a,b), the typical forms of the calculated one-dimensional functions *NL(ω)* and their asymptotes are shown.

Table 1. Asymptotes of the NLT at the first time step of KE solution.

| Run No | Initial spectrum shape | | | Asymptote |
|---|---|---|---|---|
| | *n* | *γ* | $\Psi(\theta)$ | *p* |
| 1 | 6 | 3.3 | const | 4.9 |
| 2 | 6 | 3.3 | $\cos^2(\theta)$ | 5.3 |
| 3 | 6 | 1.0 | const | 5.2 |
| 4 | 5 | 3.3 | const | 3.3 |
| 5 | 5 | 3.3 | $\cos^2(\theta)$ | 3.8 |
| 6 | 5 | 1.0 | const | 3.8 |
| 7 | 4 | 3.3 | const | 0.85 |

Note. The powers of asymptotes are found in the frequency band $5 < \omega/\omega_p < 15$.

The main features of *NL(ω)* – an alternating two-mode shape near the spectral peak and an infinite positive high-frequency tail (figure 1(a)) – are well known (Haaselmann & Hasselmann 1985; Polnikov 1989; van Vledder 2006). Here, it was found that the high-frequency tail of the NLT is positive only when n ≥ 4; when n < 4, it is negative and poorly defined due to the weak convergence of the KI for such spectra (for theoretical details, see Geojaev & Zakharov 2017; Zakharov 2017).

In general, the features of asymptotes for *NL(ω)* can be formulated as follows:

1) The high-frequency tail of the NLT is positive when n ≥ 4 and negative when n <4;

2) Representation of *NL(ω)* in form (12) has a varying power-law of decay;

3) For values *n* ≥ 5, parameter *p* for *NL(ω)* is close to value *n*-1 (i.e., *Nl (ω) ~ ωS (ω)*), in the intermediate frequency region, $5 < \omega/\omega_p < 15$; but in the entire frequency band, $5 < \omega/\omega_p < 40$, the decay of *Nl(ω)* is somewhat weaker than *ωS (ω)*;

4) With a decrease of parameter $\gamma$ and an increase of the anisotropy parameter $a$, the NLT-decay parameter $p$ increases.

5) When the decay parameter $n$ approaches 4, the relative intensity of the NLT-tail decreases radically (Figure 2(a)); when $n \leq 4$, the decay features of NLT begin to depend significantly on the spectrum parameters: $\gamma$, $a$, relative frequency $\omega/\omega_p$ (figure 2(b)), and on the limits of the computing grid. This feature is due to the slow convergence of the KI in such a case (see numerical details in Polnikov & Uma 2014; and theoretical explanation in Zakharov 2017).

From the behaviour of the asymptote for NLT, it follows that the nonlinear interactions at high frequencies are sensitive to both the frequency-shape of the spectral peak (i.e., variations of parameter $\gamma$) and the angular shape of the spectrum (i.e., variations of parameter $a$). In other words, *the nonlinear interactions have an explicit non-local feature.*

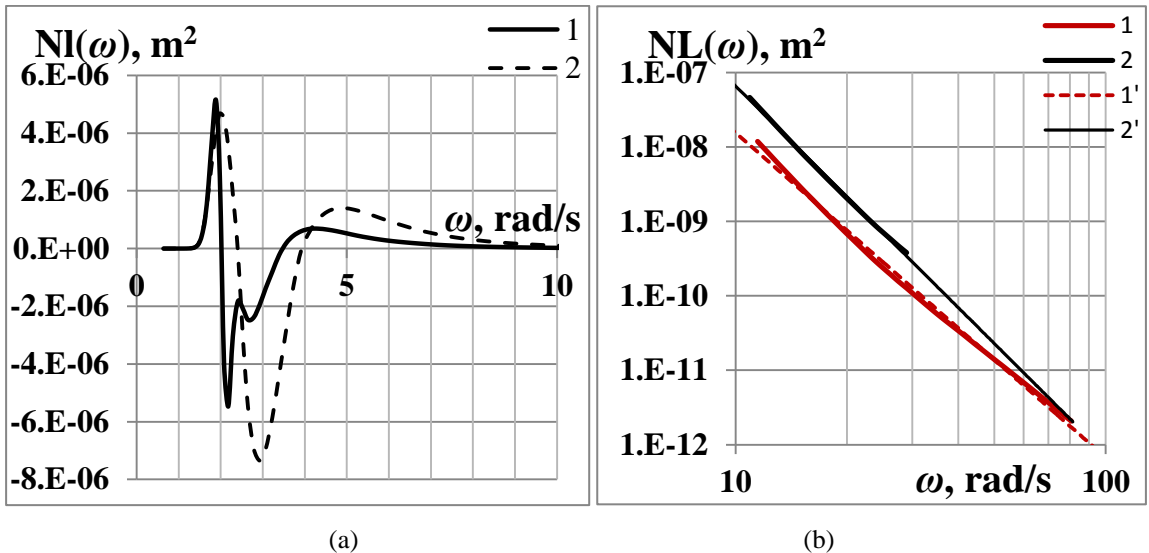

(a)               (b)

FIGURE 1. (a) Form of one-dimensional NLT, $NL(\omega)$, at low frequencies.
Line 1 corresponds to run 1 from table 1, line 2 to run 3 from table 1.
(b) Form of $NL(\omega)$ at high frequencies, run 1 from table 1. Line 1 corresponds to the whole tail of NLT, line 2 to the part of tail ($10 < \omega < 30$) rad/s (with a weight of 3, to separate the lines).
Line 1' is the root-mean-square trend of the tail-part 1 (equation, $y = 0.0004x^{-4.4}$), and
line 2' is the same for the tail-part 2 (equation, $y = 0.002x^{-4.9}$).

This conclusion had already been noted (Polnikov 1989); however, the non-locality of the four-wave nonlinear interactions is expressed in the asymptote of NLT in the clearest manner. As shown below, the non-locality of NLT is very important in explaining the reasons of the self-similar spectral shape formation.

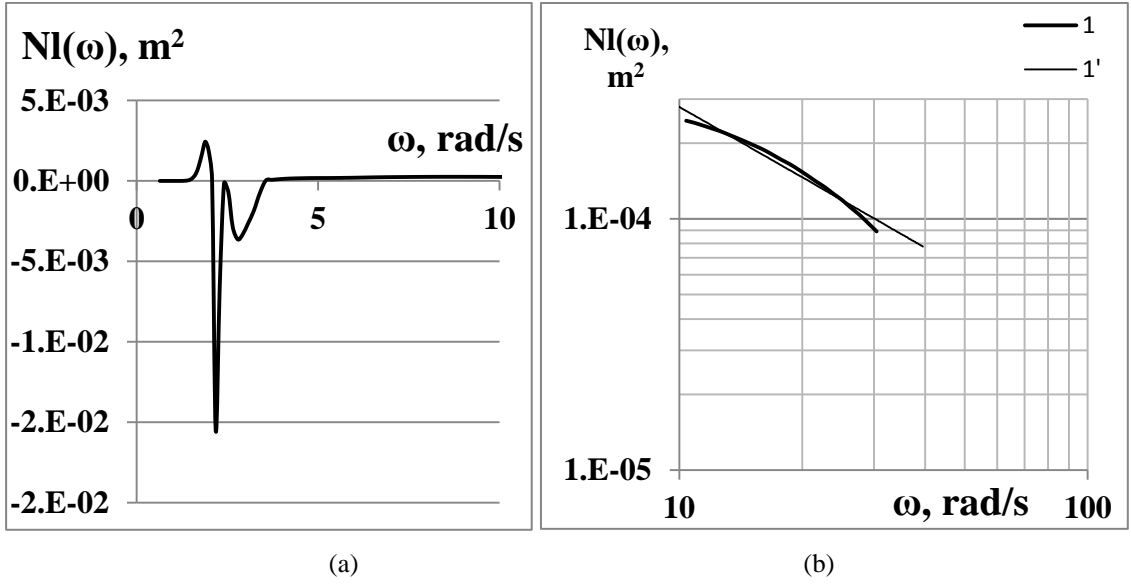

FIGURE 2. (a) Form of *NL(ω)* at low frequencies for the initial spectrum of run 5 from table 1.
(b) Shape of *NL(ω)* at high frequencies for the same spectrum. Line 1 is NLT, line 1' is the root-mean-square trend for the part of tail in the band: (10 <ω <30) rad/s (equation, y = 0.0024x$^{-0.9}$).

*3b. Kinetic equation solutions on a large time scale*

To achieve the goals posed, we performed a large series of numerical solutions for the KE of type (1) on grid (6) for the algorithm versions both with conservation of total wave action *N* and with retaining total wave energy *E*. The difference between these versions was realised by making the exact balance of positive and negative NLT for either *N* or *E*, respectively, at each time step of the KE numerical solution. For the first time, our calculations have shown that on large scales of evolution time, this difference affects neither the feature of NLT-asymptotes nor the steady-state shape of the self-similar one-dimensional spectrum, $S_{sf}(\omega)$.

For each version of the KE-solution, the following results were established:
1) The self-similar one-dimensional spectrum, $S_{sf}(\omega)$, decays with the law

$$S_{sf}(\omega) \sim \omega^{-4 \pm 0.02}; \qquad (15)$$

2) The asymptote of *Nl(ω)* is invariant in time, negative in magnitude, and takes the form

$$Nl(\omega) \sim -\omega^{-4.15 \pm 0.05}; \qquad (16)$$

3) The shape parameters for the two-dimensional spectrum $S_{sf}(\omega, \theta)$, $A_p = A(\omega_p)$ and *B*, vary slightly with time within ±5%, and their average values certainly depend on the degree of anisotropy for the initial spectrum $S(\omega,\theta)$. [2]

---

[2] Hereafter, all estimates of the powers are obtained in the EXCEL-shell with the installed trend-formulas (obtained by the method of least squares). Scattering the exponent powers, in (15), (16) and further, are due to the scattering of the obtained set of estimates, exceeding the statistical errors of the exponents.

The summarised results of calculations for a representative series of initial spectrum shapes $S(\omega,\theta)$ are presented in table 2, together with the asymptote parameters for $Nl(\omega)$ and parameters $A_p$ and $B$, for the steady-state shape, $S_{sf}(\omega, \theta)$.

Table 2. Asymptote of $Nl(\omega)$ and shape parameters of $S_{sf}(\omega,\theta)$ in the long-term KE solution.

| Run No | Initial spectrum shape | | | Evolution time, s | Asymp. of $Nl(\omega)$ | Parameters of $S_{sf}(\omega,\theta)$ | |
|---|---|---|---|---|---|---|---|
| | $n$ | $\gamma$ | $\Psi(\theta)$ | | $p$ | $B*100$ | $A_p*100$ |
| 1 | 6 | 3.3 | const | $1.3 \cdot 10^6$ | 4.1 | 22 | 16 |
| 2 | 6 | 1.0 | const | $4.2 \cdot 10^6$ | 4.2 | 25 | 16 |
| 3 | 5 | 3.3 | const | $1.3 \cdot 10^5$ | 4.1 | 25 | 16 |
| 4 | 5 | 1.0 | const | $7.9 \cdot 10^5$ | 4.2 | 25 | 16 |
| 5 | 4 | 3.3 | const | $6.9 \cdot 10^4$ | 4.1 | 23 | 16 |
| 6 | 4 | 1.0 | const | $5.4 \cdot 10^4$ | 4.2 | 26 | 16 |
| 7 | 5 | 1.0 | $\cos^2(\theta/2)$ | $4.1 \cdot 10^6$ | 4.2 | 32 | 46 |
| 8 | 5 | 1.0 | $\cos^8(\theta/2)$ | $3.6 \cdot 10^6$ | 4.1 | 34 | 63 |
| 9 | 5 | 1.0 | $\cos^2(\theta)$ | $4.1 \cdot 10^5$ | 4.2 | 33 | 64 |
| 10 | 5 | 1.0 | $\cos^4(\theta)$ | $8.7 \cdot 10^6$ | 4.2 | 33 | 66 |
| 11 | 5 | 3.3 | $\cos^{12}(\theta)$ | $5.6 \cdot 10^6$ | 4.2 | 31 | 61 |

Note. Different degrees of shadowing show differences in angular forms for self-similar spectra.
Value $B=0.33$ corresponds to $\gamma=3.3$, $B=0.63$ to $\gamma=1.0$ and $A_p= 0.16$ to $\Psi=$ const, $A_p= 0.64$ to $\Psi= \cos^2(\theta)$.

Table 2 shows that the values of $A_p$ and $B$ differ significantly in the cases of isotropic and anisotropic initial spectrum $S(\omega,\theta)$, whilst they have intermediate values (run 7) for weakly anisotropic initial spectra. For anisotropic spectra, the function of angular directivity, $A(\omega)$, includes a sharp peak slightly below the peak frequency, $\omega_p$. As the frequency increases, $A(\omega)$ quickly becomes constant, $A \approx 0.16$ (when $\omega \geq 2\omega_p$), corresponding to the isotropic distribution of the spectrum (figure 3a). In turn, the frequency shape of the one-dimensional spectrum, $S_{sf}(\omega)$,

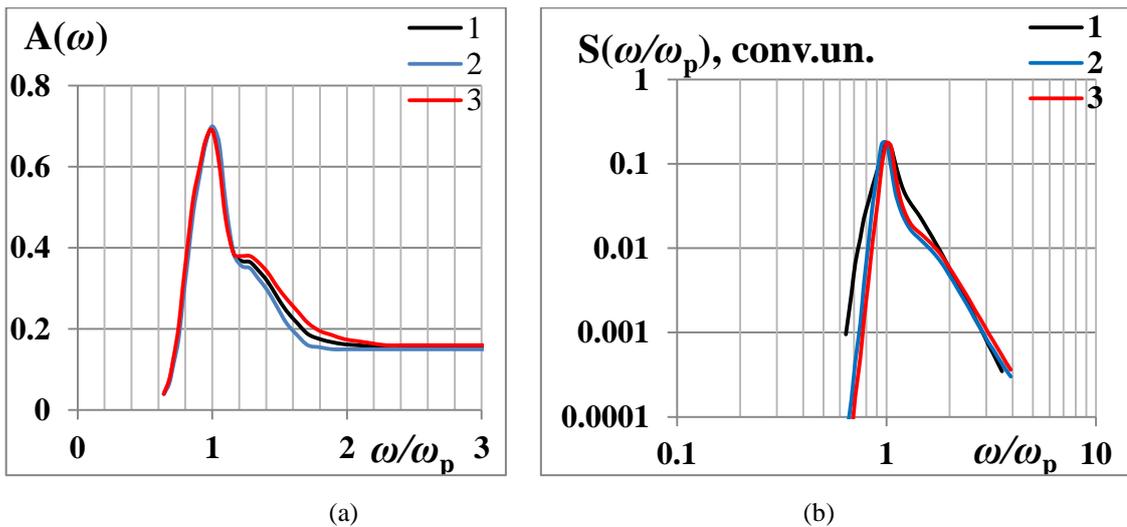

(a) (b)
FIGURE 3. (a) Typical shape of the self-similar function for the angular directivity, $A(\omega)$.
Line 1 is the mean for all anisotropic runs (7 through 11) from table 2; line 2 is run 10; line 3 is run 11.
(b) Self-similar shapes $S_{sf}(\omega/\omega_p)$ on the example of run 3 from table 2 (isotropic spectrum).
Line 1 is the shape of J-spectrum according to (7,8) for $n = 5$;
line 2 is $S_{sf}(\omega/\omega_p)$ at time $t \approx 1 \cdot 10^4$ s; line 3 is $S_{sf}(\omega/\omega_p)$ at the time $t =1 \cdot 10^5$ s.

also has a sharp peak. For isotropic initial spectra, this shape is very similar to one for the J-spectrum with parameters $\gamma = 3.3$ and $n = 5$ (figure 3b). For the anisotropic initial spectra, the shape of $S_{sf}(\omega)$ is similar to that above, although the peak is 1.5 times wider (for a value of B). Such a detailed description of the integral parameters for two-dimensional self-similar spectrum $S_{sf}(\omega, \theta)$ essentially supplements the results of earlier works (Polnikov 1990; Pushkarev et al. 2003; Badulin et al. 2005). It has a reasonable academic interest.

Detailed information regarding the features for spectrum evolution is shown in figures 4 through 7. The time-history of a one-dimensional spectrum $S(\omega)$ for run 1 from table 2 is presented in figure 4a. The corresponding time-history of one-dimensional NLP $Nl(\omega)$ is shown in figure 4b. It can be seen that when the self-similar shape of the spectrum is formed (at a time of the order of $4 \cdot 10^3$ s, or more), the NLP changes principally in such a way that the intensity of the tail of $Nl(\omega)$ becomes extremely small. Its asymptotic behaviour is given by formula (14). The same situation is realised for all the cases considered (with small variations for parameter $p$, as shown in table 2).

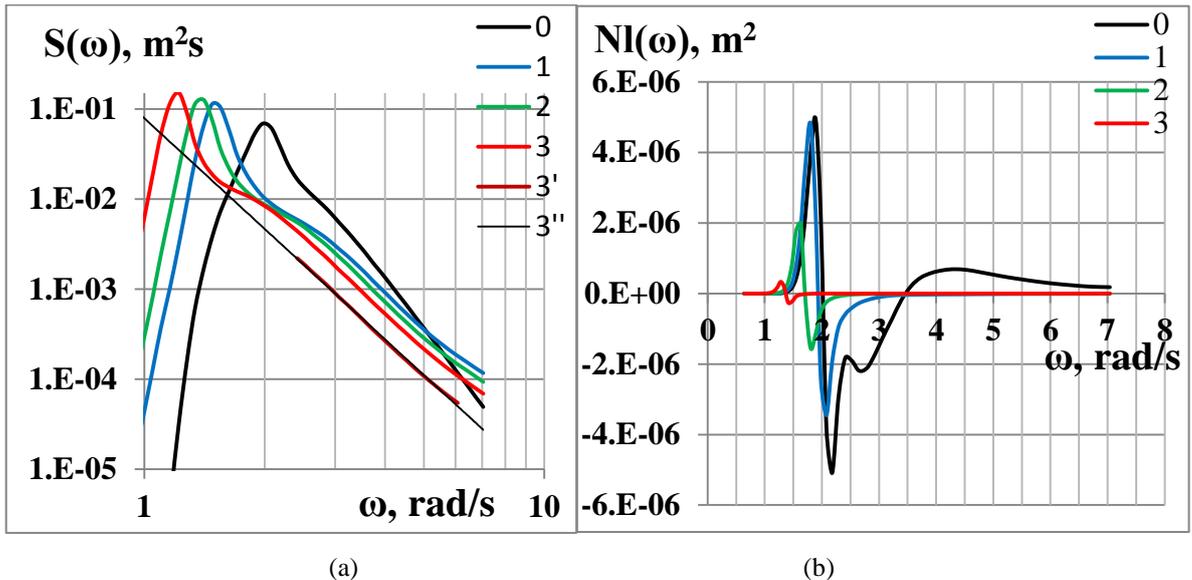

(a) (b)
FIGURE 4. (a) Time-history of one-dimensional spectrum $S(\omega, t)$ for run 1 from table 2.
Line 0 corresponds to $t = 0$ s; line 1 to $t = 1.4 \cdot 10^5$ s; line 2 to $t = 3.0 \cdot 10^5$ s; line 3 to t = $1.2 \cdot 10^6$ s;
line 3' is the tail of the spectrum of line 3 (with a weight of 0.3); 3" is the trend of tail 3'
(equation y = $0.0792x^{-4.1}$).
(b) Time-history of one-dimensional NLT $Nl(\omega, t)$ (the same run).
Line 0 corresponds to $t = 0$ s; line 1 to $t = 470$ s; line 2 to $t = 4.5 \cdot 10^4$ s; line 3 to $t = 3.7 \cdot 10^4$ s.

In this respect, run 6 from table 2 is very indicative (figures 5(a, b)). Figure 5(a) demonstrates the evolution of the initially isotropic spectrum ($a = 0$) with parameters $\gamma = 1$ and $n = 4$. In this case, figure 5(b) shows that, although the initial spectrum is decaying with the law: $S(\omega) \sim \omega^{-4}$, the intensity of the tail for nonlinear transfer $Nl(\omega)$ is not small at the first time-step, as might be expected in accordance with the conclusions by Zakharov and Filonenko (1966).

However, on the relatively great evolution scales, $t > 10^4 \cdot 1/\omega_p(0)$, a self-similar spectrum with the tail of form (13) has already formed, with the same decay-law as the initial spectrum (i.e., $S(\omega) \sim \omega^{-4}$). However, on these scales, the nonlinear transfer gets a rapidly decaying tail of form (14); the intensity of this NLT-tail is already very small, corresponding with the theory of Zakharov and Filonenko (1966).

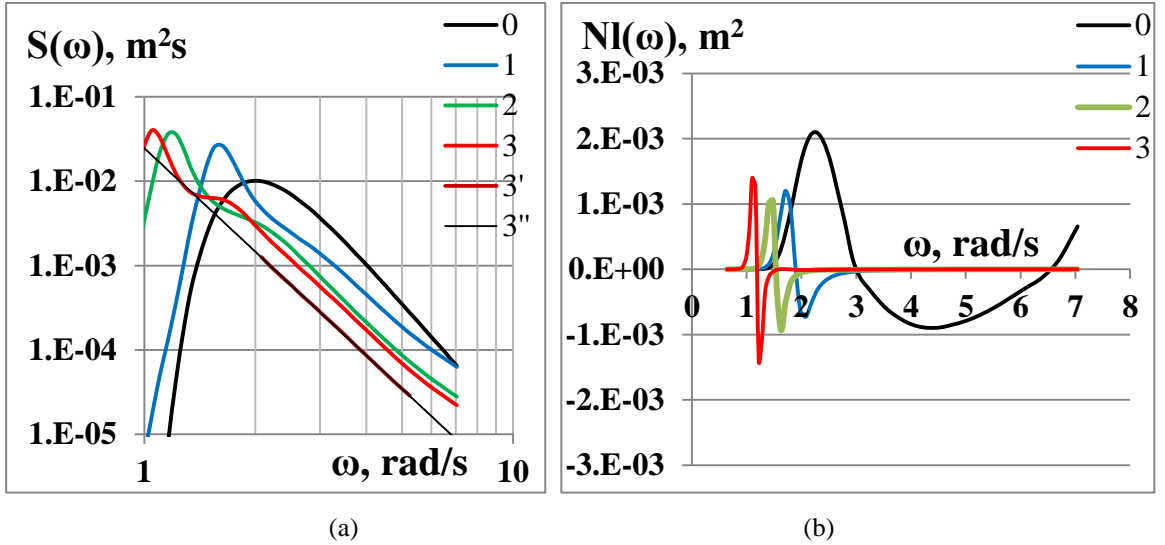

(a)            (b)

FIGURE 5. (a) Time-history of one-dimensional spectrum $S(\omega,t)$ for run 6 from table 2.
Line 0 corresponds to $t = 0$ s; line 1 to $t = 5.7 \cdot 10^2$ s; line 2 to $t = 3.9 \cdot 10^3$ s; line 3 to $t = 5.4 \cdot 10^4$ s;
line 3' is the tail of the spectrum of line 3 (with a weight of 0.3); 3" is the trend of tail 3'
(equation $y = 0.0248 x^{-4.1}$).
(b) Time-history of one-dimensional NLT $Nl(\omega, t)$ (the same run).
Line 0 corresponds to $t = 0$ s with a weight of 1); line 1 to $t = 570$ s (with a weight of 3);
line 2 to $t = 3.9 \cdot 10^3$ s (with a weight of 10); line 3 to $t = 5.4 \cdot 10^4$ s (with a weight of 100).

Undoubtedly, the main reason for the small value of the tail-intensity for function $Nl(\omega)$ at a large $t$ is the formation of the entire self-similar spectrum shape, $S_{sf}(\omega,\theta)$. However, in this case (run 6), the difference between the spectrum form $S_{sf}(\omega,\theta)$ and the initial form is determined only by the self-similar frequency-shape of the peak-domain for $S_{sf}(\omega,\theta)$, because the angular distribution of the spectrum is always isotropic. This fact testifies, first, to the nonlocality of the four-wave nonlinear interactions and, second, to the fundamental role of the spectrum-peak shape in forming the tail of the NLT-function, $Nl(\omega)$, with the frequency-asymptote of form (14).

Here, it is important to emphasise that the results presented above were repeated for versions of solutions for KE (1) both with preservation of the total wave action $N$ and with conservation of the total wave energy $E$ (see figures 8(a,b) below). Moreover, the same results were found in all cases of incomplete conservativeness for $N$ and $E$, when such a conservation control was not carried out. At the same time, some numerical results also differ for the various versions of the algorithms, as seen from the temporal asymptote for several spectral parameters, including the time-asymptote for total wave action, $N(t)$, and total wave energy, $E(t)$.

To see this, the frequency functions of fluxes for wave energy $P_E$ and wave action $P_N$ were estimated with the formulas taken from (Pushkarev et al. 2003; Badulin et al. 2005):

$$P_E(\omega) = -\int_{\omega_1}^{\omega}\int Nl(\omega,\theta)d\theta d\omega \quad \text{and} \quad P_N(\omega) = -\int_{\omega_1}^{\omega}\int (Nl(\omega,\theta)/\omega)d\theta d\omega, \tag{15}$$

where $\omega_1$ is the lower limit of the numerical frequency band. According to (Pushkarev et al. 2003; Badulin et al. 2005), a positive value of $P_E$ (or $P_N$) indicates a flux upward in frequency and a negative value downward.

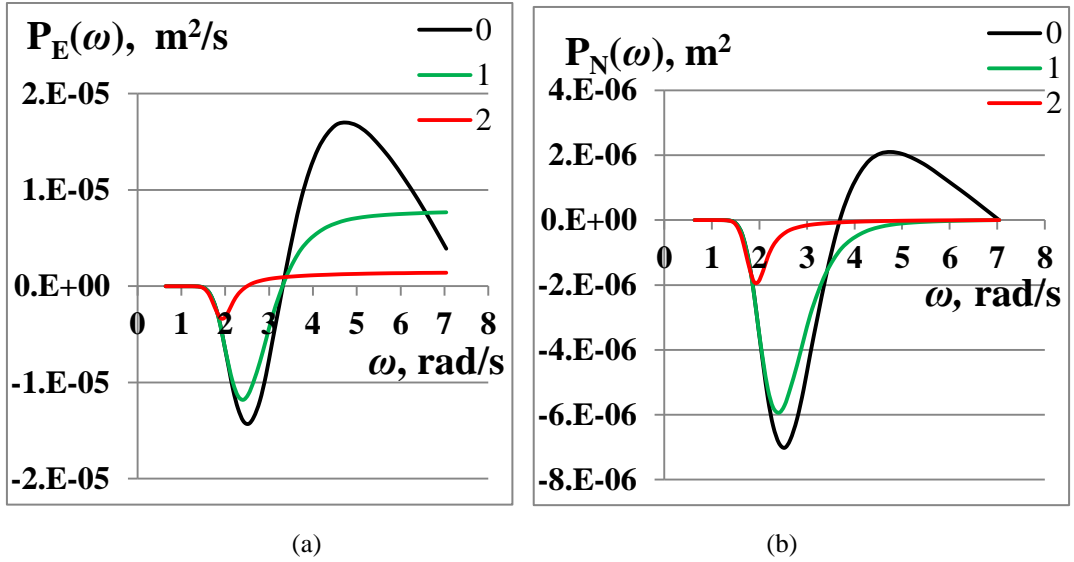

(a)          (b)

FIGURE 6. (a) Time-history of flux $P_E(\omega,t)$ for run 2 from table 2 (when $N$ is constant).
Line 0 corresponds to $t = 0$ s; line 1 to $t = 700$ s; line 2 to $t = 2500$ s.
(b) Time-history of flux $P_N(\omega,t)$ for the same run 2.
Lines correspond to the same time moments.

For run 2 from table 2, the time-history of fluxes $P_E(\omega,t)$ and $P_N(\omega,t)$ is shown in figures 6(a,b) under the condition of constant wave action $N$ and in figures 7(a,b) under the condition of constant wave energy $E$. Figures 6(a,b) show that if the wave action $N$ is constant, a constant upward energy flux $P_E$ is established in the tail frequencies range ($\omega > 2\omega_p(t)$) on a long evolution scale (t > $10^3/\omega_p(0)$). The latter means the loss of wave energy in the system. In the same frequency range, the flux $P_N(\omega,t)$ is always near zero. Under the condition of constant wave energy $E$, the opposite is true (figures 7(a, b)). Note that the frequency dependence of fluxes $P_E(\omega,t)$ and $P_N(\omega,t)$ and the temporal change in their intensities, governed by equation (15), directly follow from the time-history for functions $Nl(\omega,t)$, shown, for example, in figures 4(b) and 5(b). Note also that the shape of the self-similar spectra obtained in both of these versions of the KE- solution is absolutely the same, with differences only in intensity (figures 8(a,b)).

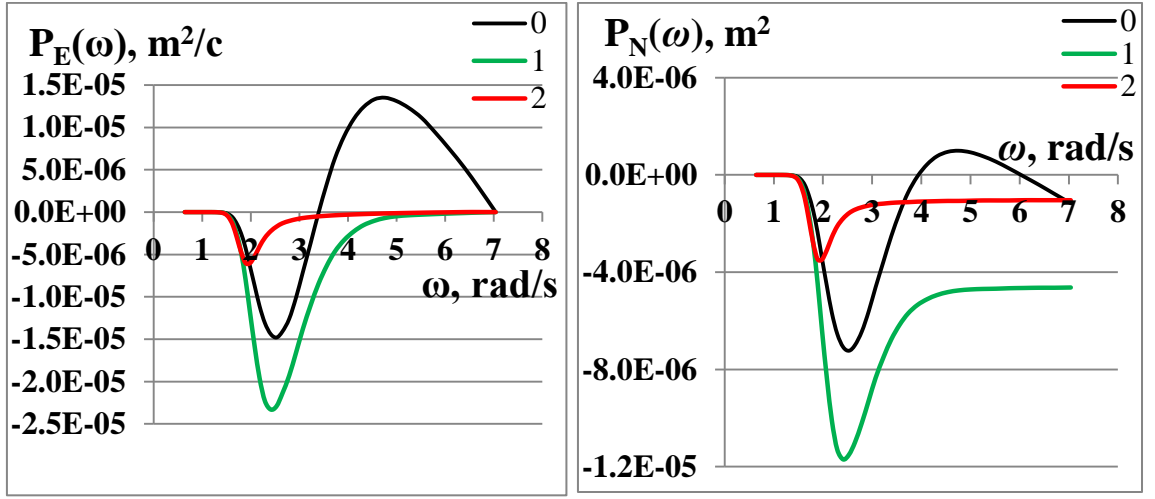

FIGURE 7. (a) Time-history of flux $P_E(\omega,t)$ for run 2 from table 2 (when $E$ is constant).
Line 0 corresponds to $t = 0$ s; line 1 to $t = 250$ s; line 2 to $t = 1830$ s.
(b) Time-history of flux $P_N(\omega,t)$ for the same run 2.
Lines correspond to the same time moments.

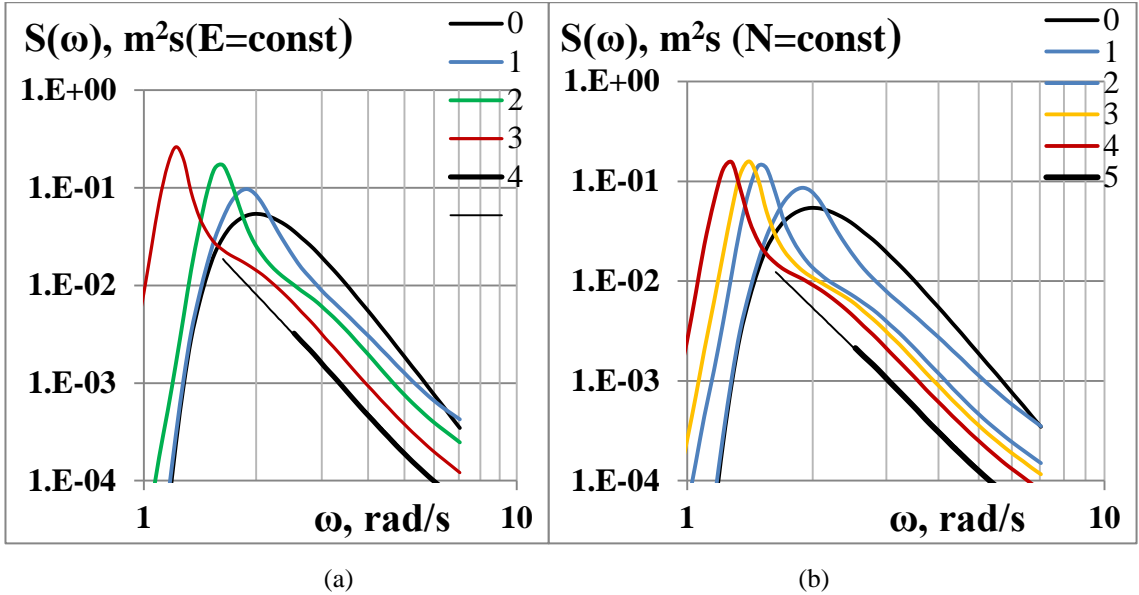

FIGURE 8. Time-history $S(\omega,t)$ for two versions of the KE-solution (run 2):
(a) the version with condition $E$ = const;
Line 0 corresponds to $t = 0$ s; line 1 to $t = 1830$ s; line 2 to $t = 0.5 \cdot 10^5$ s, line 3 to $t = 3.5 \cdot 10^5$ s,
line 4 to the trend for line 3 (with a weight of 0.5, equation y= 0,135·x$^{-4.06}$).
(b) the version with condition $N$ = const.
Line 0 corresponds to $t = 0$ s; line 1 to $t = 2500$ s; line 2 to $t = 6.7 \cdot 10^4$ s, line 3 to $t = 1.7 \cdot 10^5$ s,
line 4 to $t = 6.1 \cdot 10^5$ s line 5 to the trend for line 4 (with a weight of 0.5, equation y= 0,09·x$^{-4.06}$)

Here it is important to note that, despite the invariance of the shape of spectrum $S_{sf}(\omega,\theta)$, the process of establishing a self-similar spectrum differs somewhat for different versions of the KE-solution algorithms. These differences manifest via the temporal asymptote for the peak frequency, $\omega_p(t)$, the values of the self-similar spectrum peak, $S_p(t) = S_{sf}(\omega_p,\theta)$, and, naturally, the temporal asymptote of integral characteristics $E(t)$ and $N(t)$ (in proper case). In the asymptotes mentioned, the basic one is the temporal behaviour of the peak frequency, $\omega_p(t)$. Using the self-similarity condition for the form $S_{sf}(\omega,\theta)$, one can analytically show the following (Pushkarev et al. 2003; Badulin et al. 2005): for constant $E$, it is true that

$$\omega_p(t) \propto t^{-1/9}, \qquad (16)$$

and for constant $N$, it should be

$$\omega_p(t) \propto t^{-1/11}. \qquad (17)$$

Details of derivations for ratios (16, 17) can be found in the Appendix to this paper. But here we consider manifestations of the mentioned asymptotes on the example of our results for isotropic initial spectra.

Indeed, the self-similar one-dimensional spectrum, $S_{sf}(\omega)$, can be represented as

$$S_{sf}(\omega) \equiv \cdot S(\omega_p, t) F_{sf}(\omega/\omega_p),$$

where $F_{sf}(\omega/\omega_p)$ is the one-dimensional shape-function, independent of time. Then, the conservation of the total wave energy $E$ has the form

$$E(t) = \int S_{sf}(\omega,t) d\omega = \int S_{sf}(\omega_p,t) F_{sf}(\omega/\omega_p) d\omega \propto S_{sf}(\omega_p,t) \cdot \omega_p(t) = \text{const.} \qquad (18)$$

Consequently, the peak of the spectrum, $S_p(t) = S_{sf}(\omega_p,\theta,t)$, should grow as

$$S_p(t) \sim 1/\omega_p(t), \qquad (18a)$$

and the total wave action should grow at the same rate:

$$N(t) = E(t)/\omega_p(t) \propto 1/\omega_p(t), \qquad (19)$$

as far as $E = \text{const}$.

Our calculations show that under the condition of constant wave energy, the peak frequency of the self-similar spectrum decreases according to the law

$$\omega_p(t) \sim t^{-0.09 \pm 0.02}. \qquad (20)$$

Within the limits of power scattering, ratio (20) corresponds to ratio (16). Herewith, the asymptote of the growing intensity of the spectrum peak, $S_p(t)$, and the total wave action, $N(t)$, are as follows

$$S_p(t) \sim t^{0.08 \pm 0.02} \quad \text{and} \quad N(t) \sim t^{0.08 \pm 0.02}, \qquad (21)$$

which correspond very well to the basic dependence (20) and ratios (18, 19).

Under the condition of constant wave action $N$, ratio (18) should be replaced by the following

$$N(t) = \int \left[ S_{sf}(\omega,t)/\omega \right] d\omega \sim E(t)/\omega_p(t) = \text{const.} \qquad (22a)$$

Hence, using the above ratios, we find that the total energy, $E(t)$, should decrease as

$$E(t) \sim \omega_p(t), \qquad (22b)$$

and that, according to ratios in the left-hand side of (18), it must be satisfied the ratio

$$S_p(t) = \text{const.} \tag{22c}$$

Under the condition of constant wave action $N$, our calculations give the dependence

$$\omega_p(t) \sim t^{-0.075 \pm 0.02}. \tag{23}$$

At first sight, it is noticeably weaker than the theoretical dependence (17); however, within the limits of scattering, ratio (23) is close to (17). In this case, our calculations show that

$$E(t) \sim \omega_p(t) \sim t^{-0.075 \pm 0.02}, \tag{24a}$$

and

$$S_p(t) = \text{const.} \tag{24b}$$

This means that the expected dependences (22b, c) are also fulfilled in our calculations.

According to theoretical estimates (18a) and (22c), the only visually observable effect of the differences in these versions of the KE-calculation algorithms is the different intensity of the spectrum peak $S_p(t)$. In formulas it is seen from 21, 24b; visually, it is seen in figures 8(a,b).

In addition, it is interesting to give estimates of the self-similar spectrum values at any fixed frequency $\omega_{fix}$ in the tail region. They are determined by the ratios

$$S_{sf}(\omega_{fix}) = S_{sf}(\omega_p) \cdot F_{sf}(\omega_{fix}/\omega_p) \propto S_{sf}(\omega_p) \cdot (\omega_{fix}/\omega_p)^{-4}, \tag{25}$$

where the tail for self-similar non-dimensional shape-function $F_{sf}(\omega/\omega_p)$ is explicitly written. In the case of constant $E$, from (25) and (18a), the following dependence should be derived

$$S_{sf}(\omega_{fix}) \propto (\omega_p)^{-1} \omega_{fix}^{-4} \omega_p^4 \propto \omega_p^3. \tag{26}$$

For constant $N$, according to (22c): $S_p(t) = \text{const}$. Thus, it must be valid

$$S_{sf}(\omega_{fix}) \propto \omega_p^4. \tag{27}$$

Our calculations give the following estimates. For constant $E$, we have

$$S_{sf}(\omega_{fix}) \sim t^{-0.24 \pm 0.01}, \tag{28}$$

and for constant $N$, it reads

$$S_{sf}(\omega_{fix}) \sim t^{-0.29 \pm 0.01}. \tag{29}$$

Dependences (28, 29) correspond well with the basic numerical dependences (20) and (23) for $\omega_p(t)$, if one accounts for the above scattering in the estimates of powers. Hereby, these estimates complete the description of new peculiarities for the numerical solution of KE, established in this work.

### 4. Interpretation of results

The peculiarities of the frequency asymptote of NLT, $Nl(\omega)$, and the features of the self-similar spectrum formation (Sections 3a,b), which are accompanied by the peak-frequency asymptotes (20) and (23) and related with asymptotes (21, 28, 29), do not require special treatment. The only principal point is the search for reasons of establishing the self-similar spectrum, $S_{sf}(\omega)$, with the tail of form (13), which is realised for any version of the algorithm for solution of KE (1). Although the numerical result, $S_{sf}(\omega) \sim \omega^{-4}$, was obtained earlier under the condition of wave-action constancy (Pushkarev et al. 2003; Badulin et al. 2005), here the question arises as to why the one-dimensional self-similar spectrum, $S_{sf}(\omega)$, always has a tail of form (13), both in cases of preservation of the total wave action and in cases of conservation of the total wave energy. An interpretation of this result is given below.

First, however, we note that the possibility of separating these algorithms is stipulated by the fact that two integral parameters of the wave-system, $N$ and $E$, are not simultaneously conserved in the numerical solution of KE (1). This statement was repeatedly noted by Zakharov and co-authors (see Zakharov & Pushkarev 1999; Pushkarev et al. 2003; Badulin et al. 2005; Zakharov 2017, and references therein). In our opinion, the absence of the simultaneous preservation of $N$ and $E$ in the calculations is due to purely mathematical reasons, including a weak convergence of the KI for a slow-falling spectrum; a restricted numerical grid; systematic errors in the calculation of KI, provided by approximate estimations of contributions from the singular points in the integrand (Polnikov 1989, van Vledder 2006); and insufficient resolution. These reasons are not due to the physics of the phenomenon, because KE of form (1) has no external sources or sinks, which could violate the constancy of $N$ or $E$. Therefore, from the viewpoint of mathematics and physics for the system under study, both versions of the computing algorithm for KE (1) have a reason to exist. Both of these versions were realised in our calculations.

As shown above, the differences among these versions of the KE solutions lead to fundamental differences in the formation of fluxes for wave action $N$ and wave energy $E$. When $N$ is constant, on a scale of long-term evolution, a large frequency-range of constant upward energy flux $P_E(\omega)$ is realised in the tail range of spectrum. In turn, if $E$ is constant, the similar range of constant downward action flux $P_N(\omega)$ occurs (figures 6(a) and 7(b)). Thus, two fundamentally different physical situations arise. Nevertheless, the tail of one-dimensional self-similar spectrum, $S_{sf}(\omega,)$, always takes form (13) (figures 8(a,b)). The interpretation of this result requires a separate discussion.

Indeed, according to Zakharov and Zaslavskii (1982), the spectra of forms (3) and (13) are treated as the Kolmogorov's spectra of constant energy flux $P_E(\omega)$ directed upward in frequencies. Therefore, they must be formed only in the version of the KE solution with the condition of total wave action preservation, when proper flux $P_E(\omega)$ occurs. According to this logic, in the case of the KE solution with the condition of constant wave energy, a spectrum with the tail of form (4) should be formed, that is, the tail should take the form $S_{sf}(\omega) \sim \omega^{-11/3}$. However, such a spectrum is not observed in our calculations with the version of constant *E*. This result is not related to the accuracy of the calculations, because the numerical errors of our calculations are only 3% to 5%, whilst the error in estimating the exponent is not more than 1% to 2%. That gives grounds for distinguishing the '-4' and '-11/3' decay laws. Therefore, the numerical result of invariability of the solution of form (13) should be considered quite reliable.

Our treatment of the matter under discussion is based on the results of estimating the high-frequency asymptote for nonlinear energy transfer $Nl(\omega)$, obtained above. Indeed, by establishing a standard initial spectrum of waves in form (7, 8) with the decay index n > 4, we find that, for such *n*, the tail of $Nl(\omega)$ is always positive and decays rapidly (see figures 1(a), 2(a) and the description of asymptotes). Consequently, in the course of spectrum evolution in terms of KE (1), at the beginning stage, the intensity of the spectrum tail increases, resulting in decreasing the current power of its decay. On the time scales when the tail of the spectrum approaches to form $S(\omega)\sim\omega^{-4}$, the asymptote of $Nl(\omega)$ has already changed its sign and took form (14): $Nl(\omega) \sim -\omega^{-4}$. Mathematically, it is evident that this change of $Nl(\omega)$ stabilises the spectral shape.

It should be noted here that, due to the nonlocality of four-wave nonlinear interactions, the asymptote of $Nl(\omega)$ is determined not only by the asymptotic behaviour of the spectrum tail but also by the shape of its peak. The latter is clearly seen from the results of the calculations for run 6 from table 2, when the initial spectrum is falling with the '-4' law (see figures 5(a,b)). This manifests the role of the spectrum-peak domain in formation of the entire shape of self-similar spectrum $S_{sf}(\omega,\theta)$.

Thus, due to the reforming shapes of the spectrum-peak domain and the spectrum tail, a correspondence (coincidence) between the forms of asymptotes for $S(\omega)$ and $Nl(\omega)$ is realised, which stabilises the shape of the entire spectrum, providing the tail of form (13). Moreover, the experiments for initial spectra with n < 4 showed that, although the KI is not definite for such spectra, the stabilising process is even faster, because the tail $Nl(\omega)$ has a negative sign from the very beginning of spectrum evolution. Due to the non-locality of nonlinear interactions, the change in the spectral peak results in radical changes in the convergence of the KI, improving it. Eventually, on the scale of long-term evolution, there is no problem with the convergence of the KI (see figures 4(b), 5(b)).

Basing on the said, we assume that the establishment of the universal form of the self-similar spectrum is due only to the mathematical properties of KI itself. This property has such a nature that, despite of solution algorithm, the spectrum's self-similar shape is restored very quickly for initial spectra with any decay law, resulting in the described form for $S_{sf}(\omega,\theta)$ with tail (13). This new mathematical fact indicates a high degree of self-organisation of the nonlinear processes under consideration.

This allows us to conclude that the formation of a self-similar spectrum with tail $S(\omega) \sim \omega^{-4}$ is not connected with the constant energy or action fluxes that are formally realised in the numerical solution for KE of form (1). Nevertheless, these fluxes are present. What is the reason for their existence

In view of the absence of any external sources and sinks, we can postulate that the self-similar frequency-shape of spectral peak, located in domain $\omega_p \leq \omega \leq 2\omega_p$, plays the proper role of an internal sink or source in this system, whilst the spectrum tail range, $\omega \geq 2\omega_p$, plays the counterpartying role. Herewith, in the upper frequency range, $\omega \geq 2\omega_p$, the peak-domain is followed by a sharp transition in angular function $\Psi(\omega,\theta)$, resulting in an isotropically distributed power-law tail (13) (see figure 3(a)). Eventually, the self-similar spectrum with the tail of form $S(\omega) \sim \omega^{-4}$ is provided only by the mathematical properties of KI, which were established and proved analytically by Zakharov and Filonenko (1966) (in the part of principal possibility of existence of spectral solutions of form $S(\omega) \sim \omega^{-4}$).

Summarising this interpretation of the obtained results, one can state that in this problem we are not dealing with Kolmogorov's turbulence. Such a conclusion is justified for three reasons: a) the presence of a distinguished frequency scale in the system (provided by the peak of the spectrum); b) the nonlocality of four-wave nonlinear interactions formatting the entire spectrum; and c) the absence of sources and sinks external to the waves. In this case, the observed constant fluxes $P_E(\omega)$ and $P_N(\omega)$ and the leakage of total energy $E$ with asymptote (24a) or the growth of action $N$ in form (21) are merely the formal mathematical consequences of the accepted versions of the computational algorithm for solving KE (1). As far as the final spectrum-tail shape is independent of fluxes, these fluxes are not the reason of establishing the self-similar spectrum with the tail of form (13).

In addition, it is pertinent to note here that, in this problem, using the condition for wave energy constancy is preferable, from the physical point of view. It is because energy $E$ is a real physical quantity, whilst action $N$ is only an auxiliary analytical variable, the invariance of which is physically unjustified. According to Zakharov (1968, 1999, 2017), the wave action has a meaning of virtual "wave particles". Thus, the formal fact of the increase in $N$, in the case of constant energy of the system, could be interpreted as the 'condensation' of wave-particles from the tail of the spectrum.

A completely different situation of Kolmogorov's turbulence is realised under the condition of separation between source and sink, which are external to the waves, as modelled by (Polnikov 1990, 1994; Pushkarev et al. 2003; Badulin et al. 2005; and so on) in the problem of numerical solution of the KE of form (6). However, a discussion of the conditions for obtaining Kolmogorov's spectra of forms (3) and (4), a detailed analysis of this process, and its reconciliation with the conclusions of this paper requires further study and a separate presentation in future.

In addition to these prospects, it should be noted that the temporal asymptote $\omega_p(t)$ of forms (20) and (23) obtained here are somewhat weaker than the theoretical forms, (16) and (17). This is apparently due to the limited variability range for values of $\omega_p$ in our calculations, determined by grid (8), and to the relatively small scales of the considered evolution time for the spectrum. In prospect, this point requires additional study, including the cases of evolution of the spectra, which initially have anisotropic angles.

**Conclusions**

Summarising the results above, we can draw the following conclusions.

First, the features of the high-frequency asymptotes for nonlinear energy transfer $Nl(\omega)$, which are close to a power-law form, play a fundamentally important role in our understanding of the evolution of nonlinear gravity waves in water, governed by the four-wave KE of form (1). In particular, the behaviour of these asymptotes indicate the fact of nonlocality of nonlinear interactions under consideration. In addition, the change of sign for $Nl(\omega)$ and its asymptote to form $Nl(\omega) \sim -\omega^{-4}$, associated with a change in the decay-law of the spectrum tail to form $S(\omega) \sim \omega^{-4}$, is responsible for maintaining the tail shape of self-similar spectrum in the form $S_{sf}(\omega) \sim \omega^{-4}$.

Second, the shape of the peak-domain for two-dimensional spectrum $S(\omega,\theta)$, the characteristics of which are presented in table 2 and in figures 3(a, b), is responsible for the formation of self-similar two-dimensional spectrum $S_{sf}(\omega,\theta)$ as a whole, what is clearly visible on the example of calculations for run 6 from table 2, for the initial spectrum with $n = 4$ (figures 5(a, b)).

Third, a self-similar two-dimensional spectrum $S_{sf}(\omega,\theta)$ with the tail of form $S_{sf}(\omega) \sim \omega^{-4}$, supported by a nonlinear energy transfer with the tail of form $Nl(\omega) \sim \omega^{-4.15 \pm 0.05}$, is established in solutions of KE (1) for any initial spectral forms (7, 8), both in the regime of conservation of total wave energy $E$ and in the regime of constancy for total wave action $N$ (figures 8(a,b)). As a result, the establishing self-similar spectra with the tail $S_{sf}(\omega) \sim \omega^{-4}$ is the purely mathematical property of the four-wave KI, that was analytically found by Zakharov and Filonenko (1996).

The constant downward fluxes of wave action $P_N(\omega)$ and the upward energy fluxes $P_E(\omega)$ are not the reasons for self-similar spectrum formation in the course of solving KE (1), because the situation differs from that of Kolmogorov's one. The differences between these regimes and the Kolmogorov's turbulence are: a) the presence of a distinguished frequency scale in the system (provided by the peak of spectrum), b) the nonlocality of four-wave nonlinear interactions formatting the entire spectrum, and c) the absence of sources and sinks external to the waves.

Fourth, the constant fluxes, $P_E(\omega)$ and $P_N(\omega)$, realised in the calculations, and the leakage of total energy $E$ according to ratio (24a) or growing action $N$ as (21) are only the formal mathematical consequences of the accepted versions of the computational algorithm for solving KE (1), rather than a reason to establish a self-similar spectrum with the tail of form (13). In both cases, to maintain these fluxes, the peak-domain for self-similar spectrum $S_{sf}(\omega,\theta)$ formally plays the role of internal source or sink, depending on the case considered, whilst the spectrum tail plays the counterpart role.

Fifth, in calculations of KI, the conservation of total energy $E$ is preferable from the physical point of view. This preference is determined by the fact that energy is a real physical quantity, the conservation of which is laid down during the derivation of KI (Hasselmann 1962; Zakharov 1968), whilst the wave action, $N$, is only the auxiliary variable of the theory, the invariance of which is not prescribed by the initial Euler's equations. From this point of view, the previous results with Kolmogorov's spectra formation (Polnikov 1990, 1994; Pushkarev et al. 2003; Badulin et al. 2005), based on solving the extended KE of form (6), require further elaboration by means of additional numerical study of the features of the processes for Kolmogorov's spectra establishing in forms (3) and (4) both for isotropic and anisotropic initial spectra, sources and sinks.

**Acknowledgements**. The authors are grateful to participants of the session on nonlinear phenomena, have been held at the Shirshov Institute of Oceanology of RAS (December of 2017, Moscow), for their interest in our work and for their valuable criticisms, which have helped to elaborate the text of this paper. This study was carried out with the partial support of RFBR project No. 18-05-00161. The research was jointly supported by the NSFC-Shandong Joint Fund for Marine Science Research Centers under Grant U1606405.

**Appendix**

The derivation of relationships for a temporal asymptote of the peak-frequency for self-similar spectrum, $\omega_p(t)$, presented by Pushkarev et al. (2003) and Badulin et al. (2005) in forms (16-17), is insufficiently complete. For the purpose of their explanation, we present here the main analytical derivations.

First, consider the situation in which the self-similar spectrum is realised under the condition of conservation for the total wave action, $N$. As already noted in the text, in this case: a) the constant energy flux, $P_E$, is formed; and b) the peak of the spectrum is constant, $S_p \equiv S_{sf}(\omega_p, t) = $ const (see formulas 18, 19, 22). Let us write the definition for energy change, as a consequence of the nonlinear transfer (NLT) of energy:

$$dE/dt = -\frac{d}{dt}\left(\int S_{sf}(\omega)d\omega\right) = -\int (dS_{sf}(\omega)/dt)d\omega = -\int Nl(\omega)d\omega \propto (g^{-4}\omega_p^{11}S_p^3)\omega_p. \quad (A1)$$

Here, in the last proportionality ratio, the self-similarity of NLT is taken into account: $Nl_{sf}(\omega,t) = Nl_{sf}(\omega_p,t) \cdot F_{sfN}(\omega/\omega_p) \propto (g^{-4}\omega_p^{11}S_p^3) \cdot F_{sfN}(\omega/\omega_p)$, with the NLT-dimension written in parentheses (Masuda 1980; Polnikov 1989). Now, combining formula $E \sim S_p \cdot \omega_p$, the constancy of $S_p$, and the final expression in the right-hand side of (A1), one obtains the differential equation for $\omega_p(t)$ (which is valid for large values of $t$):

$$dE/dt \propto d[S_p(t)\omega_p(t)]/dt = S_p[d\omega_p(t)/dt] \propto \omega_p^{12}(t)S_p^3 \propto \omega_p^{12}(t). \quad (A2)$$

From equation (A2), the solution $\omega_p(t)$ of form (17) follows directly: $\omega_p(t) \propto t^{-1/11}$.

In the case $E = $ const, the dependence $\omega_p(t)$ is derived in a similar manner. Now, the equation for the wave-action change with time takes the form

$$dN/dt = -\frac{d}{dt}\left(\int \left[S_{sf}(\omega)/\omega\right]d\omega\right) = -\left(\int [\{dS_{sf}(\omega)/dt\}/\omega]d\omega\right) \propto g^{-4}\omega_p^{11}S_p^3, \quad (A3)$$

where the same dimensional considerations are used as in (A1), regarding the self-similarity of NLT. Furthermore, using the facts that $N \sim (S_p \cdot \omega_p)/\omega_p \sim S_p$, and $S_p \propto \omega_p^{-1}$ (see (18a) in the main text), from (A3) we obtain the proper differential equation for $\omega_p(t)$:

$$dN/dt \propto dS_p(t)/dt \propto d[\omega_p^{-1}(t)]/dt \propto \omega_p^{-2}(t)[d\omega_p(t)/dt] \propto \omega_p^{11}(t)S_p^3(t) \propto \omega_p^8(t). \quad (A4)$$

From (A4), it follows the solution of form (16): $\omega_p(t) \propto t^{-1/9}$.

It remains to note that the formulas, used to obtain $\omega_p(t)$, in no way depend on the degree of spectrum anisotropy. They are based only on the following assumptions: a) the self-similarity of both the spectrum, $S_{sf}(\omega,t)$, and NLT, $Nl_{sf}(\omega,t)$; b) the conditions for the conservation of the total wave action $N$ or wave energy $E$ (i.e., ratios (22) and (18), respectively); and c) the well-known dimension for the NLT.